# Computational screening of high-performance optoelectronic materials using OptB88vdW and TB-mBJ formalisms

Kamal Choudhary[1], Qin Zhang[2], Andrew C. E. Reid[1], Sugata Chowdhury[2], Nhan Van Nguyen[2], Zachary Trautt[1,3], Marcus W. Newrock[3], Faical Yannick Congo[1], Francesca Tavazza[1]

1 Materials Science and Engineering Division, National Institute of Standards and Technology, Gaithersburg, Maryland 20899, USA

2 Physical Measurement Laboratory, National Institute of Standards and Technology, Gaithersburg, Maryland 20899, USA

3 Office of Data and Informatics, National Institute of Standards and Technology, Gaithersburg, Maryland 20899, USA

## Abstract:

We perform high-throughput density functional theory (DFT) calculations for optoelectronic properties (electronic bandgap and frequency dependent dielectric function) using the OptB88vdW functional (OPT) and the Tran-Blaha modified Becke Johnson potential (MBJ). This data is distributed publicly through JARVIS-DFT database. We used this data to evaluate the differences between these two formalisms and quantify their accuracy, comparing to experimental data whenever applicable. At present, we have 17,805 OPT and 7,358 MBJ bandgaps and dielectric functions. MBJ is found to predict better bandgaps and dielectric functions than OPT, so it can be used to improve the well-known bandgap problem of DFT in a relatively inexpensive way. The peak positions in dielectric functions obtained with OPT and MBJ are in comparable agreement with experiments. The data is available on our websites http://www.ctcms.nist.gov/~knc6/JVASP.html and https://jarvis.nist.gov .

**Background & Summary:**

Optoelectronic properties, such as fundamental electronic bandgaps and dielectric functions, provide important material information in designing optoelectronic devices for a variety of applications, such as photovoltaic cells [1], light emitting diodes [2], transparent electronics[3], dynamic random access memory[4], astronomical devices[5], and smaller and faster devices[6]. For industrial advancement in these industries, there is a great need to synthesize cheaper, more efficient, and tunable devices. Designing these new materials requires knowledge of already available ones, which can then be tailored for a particular application. Databases dedicated to optoelectronic materials meet this need. However, such user-friendly and easy-accessible public databases are still in the development phase. Computationally, it is much easier to provide properties for thousands of materials in a systematic way than to do so through experiments. Density functional theory (DFT) is the tool of choice to compute these properties in a high-throughput manner.

It is important to note that the term 'bandgap' generally refers to the fundamental gap and not the optical gap. The difference between these quantities could be small in semiconductors but significant in insulators[7]. Materials Genome Initiative based projects such as the Materials Project (MP) [8], the open quantum materials database (OQMD)[9], and AFLOW[10] have successfully enumerated bandgaps of hundreds of thousands of materials using the generalized-gradient-approximation Perdew-Burke-Ernzerhof functional (GGA-PBE)[11] and +U corrections. MP has also calibrated the static dielectric constant of 1056 materials using density functional perturbation theory (DFPT) [12], but frequency-dependent dielectric functional data is missing. Although PBE provides great insights in distinguishing non-metallic materials, the bandgaps of materials are generally underestimated typically by 30 % to 100 %[13,14], hindering its practical application in the fields of semiconductors, photovoltaic materials, and thermoelectric devices. Other systematic databases of optoelectronic materials include Zunger et al.[15] work for photovoltaic materials using Green function screened coulomb (GW) calculations, and Castelli et al[16] work on energy-harvesting materials using the Gritsenko- Leeuwen-Lenthe-Baerends (GLLB-SC) functional. GW is much more reliable than PBE in computing optoelectronic properties. However, its high computational cost severely limits its application in high-throughput screening. Catelli's work is also limited, containing information for only about 2400 materials.

Various techniques have been used to improve bandgap prediction at a moderate computational cost, including Chan and Ceder (delta-sol)[14], modified Becke-Johnson potential [17-19], and empirical fits by Setyawan et al.[20]. Recently, the modified Becke-Johnson (MBJ) potential introduced by Tran and Blaha [17-19] has been proven to improve the bandgap description in a computationally efficient way. This potential has been successfully used in characterizing electronic properties of nonmagnetic transition-metal oxides and sulfides, metals, (anti) ferromagnetic insulators, dielectric and topological insulators[19,21-24]

In this work, we have identified a sweet spot between the computational expense and accuracy for describing optoelectronic properties by using MBJ potential in a high-throughput approach. At present, we have 7358 MBJ bandgap and frequency-dependent dielectric function entries, and the database is still growing. Additionally, we computed 17805 bandgaps and frequency-dependent dielectric functions using OptB88vdW (OPT) for comparison purposes. OPT is a Van der Waal-dispersion functional (vdW-DF) with non-local correction, which can predict crystal-structure geometry, and is essential to the calculation of optoelectronic properties, especially for anisotropic materials. The OPT functional has not only been proven to reduce error in lattice constants, but its combination with MBJ functional is known to predict bandgaps of materials[25] successfully. In addition, the error in lattice constants can significantly impact the error in optoelectronic properties such as refractive indices, and hence birefringence of non-cubic class materials. Thus, for a better description of lattice constant and bandgaps of materials, it is necessary to first optimize geometries with vdW functional such OPT. OPT is also known to predict reasonable geometrical structures for non-vdW bonded structures[26].

We validate our computational results in a few cases through comparison with experimental values. We create a public JARVIS database of our results available at https://www.ctcms.nist.gov/~knc6/JVASP.html . The data is also available in REST-API format at https://jarvis.nist.gov/ and Cloud of Reproducible Records (CoRR) at NIST (https://mgi.nist.gov/cloud-reproducible-records ). We provide the code used in this work at github page: https://github.com/usnistgov/jarvis.

**Methods:**

The methodology supporting the current work consisted of several steps, including density functional theory calculations and experimental validation of a few data points. The overall processes are shown in Fig. 1 and each step is explained in detail below.

**1 Density functional theory setup:**

The DFT calculations are performed using the Vienna Ab-initio Simulation Package (VASP)[27,28] and the projector-augmented wave (PAW) method[29]. Please note commercial software is identified to specify procedures. Such identification does not imply recommendation by the National Institute of Standards and Technology. The crystal structures were obtained from the Materials Project (MP) DFT database. More specifically, we obtained all the crystal structures with less than 30 atoms per unit cell from MP, and the potential candidates for low dimensional materials using lattice-constant criteria[30] and data-mining approaches[31]. We convert the crystal cells into its primitive cell representation before a DFT calculation. If the primitive cell and corresponding conventional cell of a crystal-structure have the same number of atoms, then we prefer conventional cell as the DFT input structure.

As the error in lattice constants can significantly impact the error in optoelectronic properties, such as refractive indices and birefringence of non-cubic class materials, we re-optimized MP geometric structures using the OPT functional[26,32]. PBE is known to report good lattice constants for materials, but its applicability to vdW-bonded materials is questionable. Recently, around 5000 materials have been proposed to be vdW-bonded using lattice-constant criteria[30] and data-mining approaches[31], signifying that a correct treatment of the vdW interactions is more important than previously thought. OPT is part of vdW-DF functional, which is a non-local correlation functional that approximately accounts for dispersion interactions. OPT has been recently determined to perform well for bulk solids as well as vdW bonded structures[26]. In a recent work by Tawfik et al.[33], OPT was proven to be one of the most accurate functionals to capture vdW interactions among several other methods. We performed plane-wave energy cut-off and k-point convergences with 0.001 eV tolerance on energy. We assumed that satisfactory energy convergence would extrapolate to reasonably converged optical property calculations as well. The structure relaxation with OPT functional was obtained with $10^{-8}$ eV energy tolerance and 0.001 eV/Å force-convergence criteria.

Next, we computed bandgap and optical properties with both OPT and MBJ in subsequent DFT calculations. In the MBJ calculations, we started from OPT-relaxed structures because the MBJ functional is a potential-only functional, which implies that we cannot compute Hellmann-Feynman forces with MBJ, hence ionic relaxations were not performed using MBJ. The OPT functional has not only been proven to reduce error in lattice constants, but its combination with MBJ functional is known to predict correct bandgaps[25] as shown for few vdW bonded materials. The MBJ potential is given by:

$$v^{mBJ}{}_{x}(r) = c v_{x}^{BR}(r) + (3c-2)\frac{1}{\pi}\sqrt{\frac{5}{12}}\sqrt{\frac{2t(r)}{\rho(r)}} \tag{1}$$

where $c$ is a system-dependent parameter, with $c$ = 1 corresponding to the Becke-Roussel (BR) potential $v_{x}^{BR}(r)$, which was originally proposed to mimic the Slater potential, the Coulomb potential corresponding to the exact exchange hole[34]. For bulk crystalline materials, Tran and Blaha proposed to determine $c$ by the following empirical relation:

$$c = \alpha + \beta\left(\frac{1}{V_{cell}}\int_{V_{cell}}\frac{|\nabla\rho(r)|}{\rho(r)}dr\right)^{1/2} \tag{2}$$

With $\alpha = -0.012$, $\beta = 0.541$ Å$^{1/2}$ and $V_{cell}$ is the volume of the unit cell. The $c$-parameter was automatically determined in VASP through a self-consistent run.

To obtain the optical properties of the materials, we calculated the imaginary part of the dielectric function from the Bloch wavefunctions and eigenvalues[35,36] (neglecting local field effects). We introduced three times as many empty conduction bands as valance bands. This treatment is necessary to facilitate proper electronic transitions. We choose 5000 frequency grid points to have a sufficiently high resolution in dielectric function spectra. The imaginary part is calculated as:

$$\varepsilon^{2}{}_{\alpha\beta}(\omega)=\frac{4\pi^2 e^2}{\Omega}\lim_{q\to 0}\frac{1}{q^2}\sum_{c,v,k}2w_{\vec{k}}\delta(\zeta_{ck}-\zeta_{vk}-\omega)\left\langle\psi_{c\vec{k}+\vec{e}_\alpha q}\middle|\psi_{v\vec{k}}\right\rangle\left\langle\psi_{c\vec{k}+\vec{e}_\beta}\middle|\psi_{v\vec{k}}\right\rangle^{*} \tag{3}$$

where $e$ is electron charge, $\Omega$ is the cell volume, $w_{\vec{k}}$ is the Fermi-weight of each $k$-point, $e_\alpha$ are unit vectors along the three Cartesian directions, $\left|\psi_{n\vec{k}}\right\rangle$ is the cell-periodic part of the pseudopotential wavefunction for band $n$ and k-point $k$, $q$ stands for the Bloch vector of an incident wave, $c$ and $v$ stand for conduction and valence bands, $\xi$ stands for eigenvalues of the corresponding bands respectively. The matrix elements on the right side of Eq. (3) capture the transitions allowed by symmetry and selection rules [37]. The real part of the dielectric tensor $\varepsilon^1$ is obtained by the usual Kramers-Kronig transformation [35]

$$\varepsilon^{1}{}_{\alpha\beta}(\omega)=1+\frac{2}{\pi}P\int_0^\infty\frac{\varepsilon^2_{\alpha\beta}(\omega')\omega'}{\omega'^2-\omega^2+i\eta}d\omega' \tag{4}$$

where $P$ denotes the principle value, and $\eta$ is the complex shift parameter taken as 0.1.

It is to be noted that in conventional DFT, excited states are not optimized, hence many-body interactions are missing. To get the excited state optical properties, a high-level calculation such as the Bethe-Salpeter equation (BSE)[38] is needed, however, the conventional DFT data remains useful for qualitative comparison.

## 2 Experimental details:

We validated our DFT dielectric function data for 2H-$MoS_2$, 1T-$SnSe_2$, Si, Ge, GaAs and InP comparing to experiments. We perform our experimental measurements for 2H-$MoS_2$, 1T-$SnSe_2$.

Other experimental data were taken from Aspnes et al.[39] for validation. 1T-$SnSe_2$ (40 nm thickness) was grown on a GaAs (111) substrate by molecular beam epitaxy (MBE)[40]. The GaAs substrate was deoxidized in-situ under ultra-high vacuum (4 x $10^{-8}$ Pa) at 690 ºC for 3 min and annealed under a flux of Se for 20 min, which provides a smoother growth surface. After the substrate was cooled down and held at the growth temperature of 200 ºC for 40 min, sixty-three layers (≈ 40 nm) of 1T-$SnSe_2$ were grown by a simultaneous incidence of Sn and Se at a rate of 1/38 layer per second based on Reflection High-Energy Electron Diffraction (RHEED) oscillations. The beam equivalent pressures (BEPs) for Sn and Se, supplied by using Knudsen cells, are 2.67 x $10^{-6}$ Pa (2 x $10^{-8}$ Torr) and 2.67 x $10^{-4}$ Pa (2 x $10^{-6}$ Torr), respectively. The single phase and high crystallinity of $SnSe_2$ were confirmed by X-ray diffraction (XRD). Bulk $MoS_2$ was commercially purchased from SPI Supplies[41]. Please note the commercial product is identified to specify procedures. Such identification does not imply recommendation by the National Institute of Standards and Technology. The dielectric functions were obtained from spectroscopic ellipsometry (SE). The SE measurements were performed in a nitrogen gas-filled chamber at room temperature on a vacuum ultraviolet (UV) spectroscopic ellipsometer with a light photon energy from (0.7 eV to 8.0) eV in steps of 0.02 eV for $SnSe_2$ and from (1.0 eV to 9.0) eV in steps of 0.01 eV for $MoS_2$, at an angle of incidence of 70°.

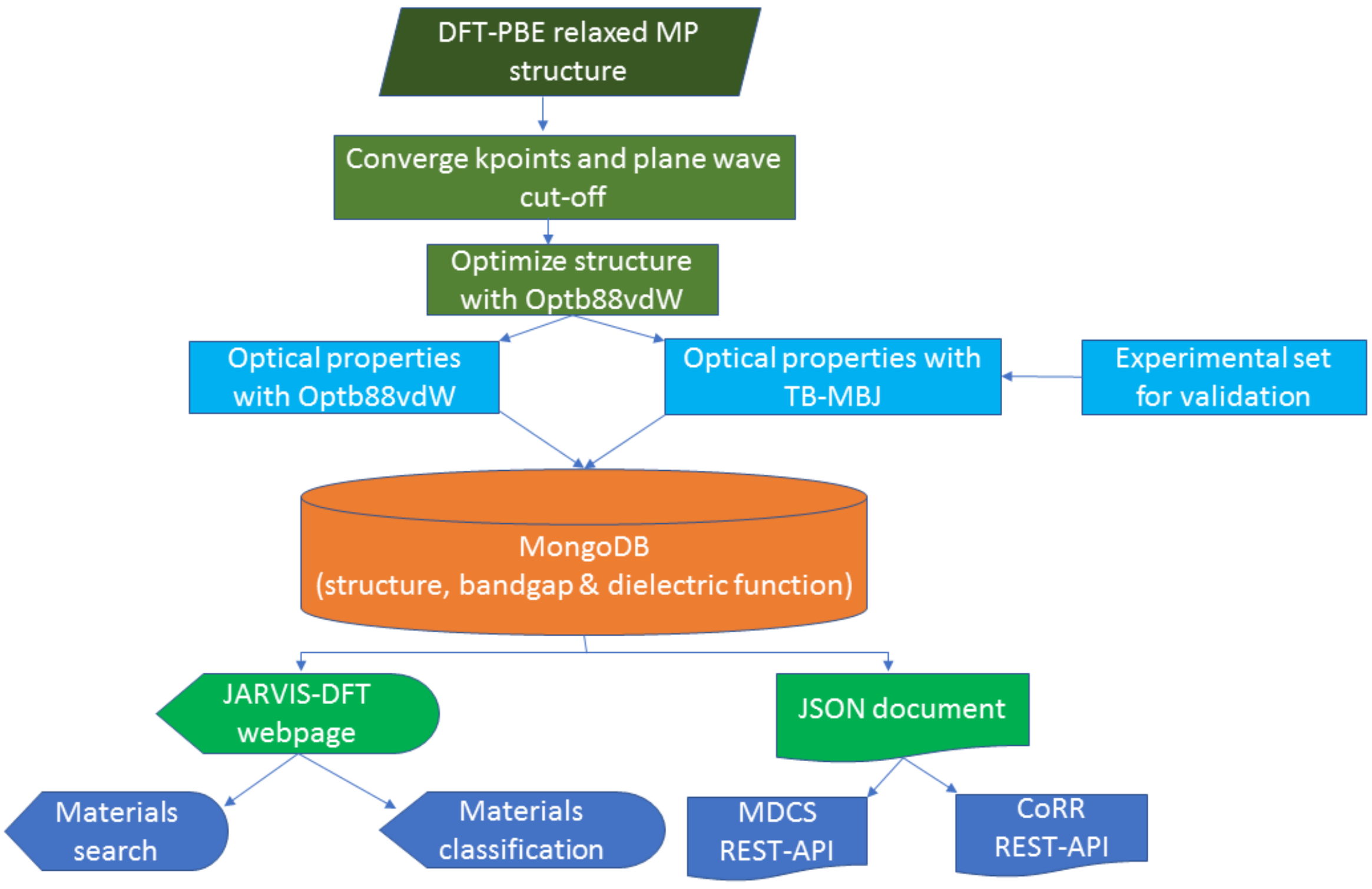

Fig. 1 Flowchart for calculating bandgap and dielectric function of materials using density functional theory.

**User-interface:**

The data is presented in a webpage format (https://www.ctcms.nist.gov/~knc6/JVASP.html ). First, a user selects the desired element/elements in the periodic table provided at the website and clicks on the 'Search' button (as shown in Fig. 2). This procedure generates a data table on the webpage consisting of the calculation-identifier, the formula of the structure, the functional used in the calculation, bandgap, mechanical property, space group of crystal and energetics of the system. Next, the user clicks on the calculation identifier for a formula, space group and functional and property data for detailed information. The detailed page is provided in the format such as

https://www.ctcms.nist.gov/~knc6/jsmol/JVASP-1174.html where ‘1174’ denotes an identifier and can assume any JARVIS-ID. The particular webpage consists first of an interactive crystal visualization, then geometric properties such as computational XRD, bandstructure and the optical properties consisting of dielectric function and refractive index. We also provide a classification of materials based on their OPT and MBJ based bandgaps, and static refractive index data as shown in Fig. 3. Clicking on one of the options in Fig. 3 results in materials with classified properties. For example, clicking on ‘Classification of 3D-bulk materials based on TB-MBJ-bandgap’ produces a table with materials that have a bandgap in rage from 0 to 1, 1 to 2, 3 to 4 eV and so on. Each material is hyperlinked to its specific webpage.

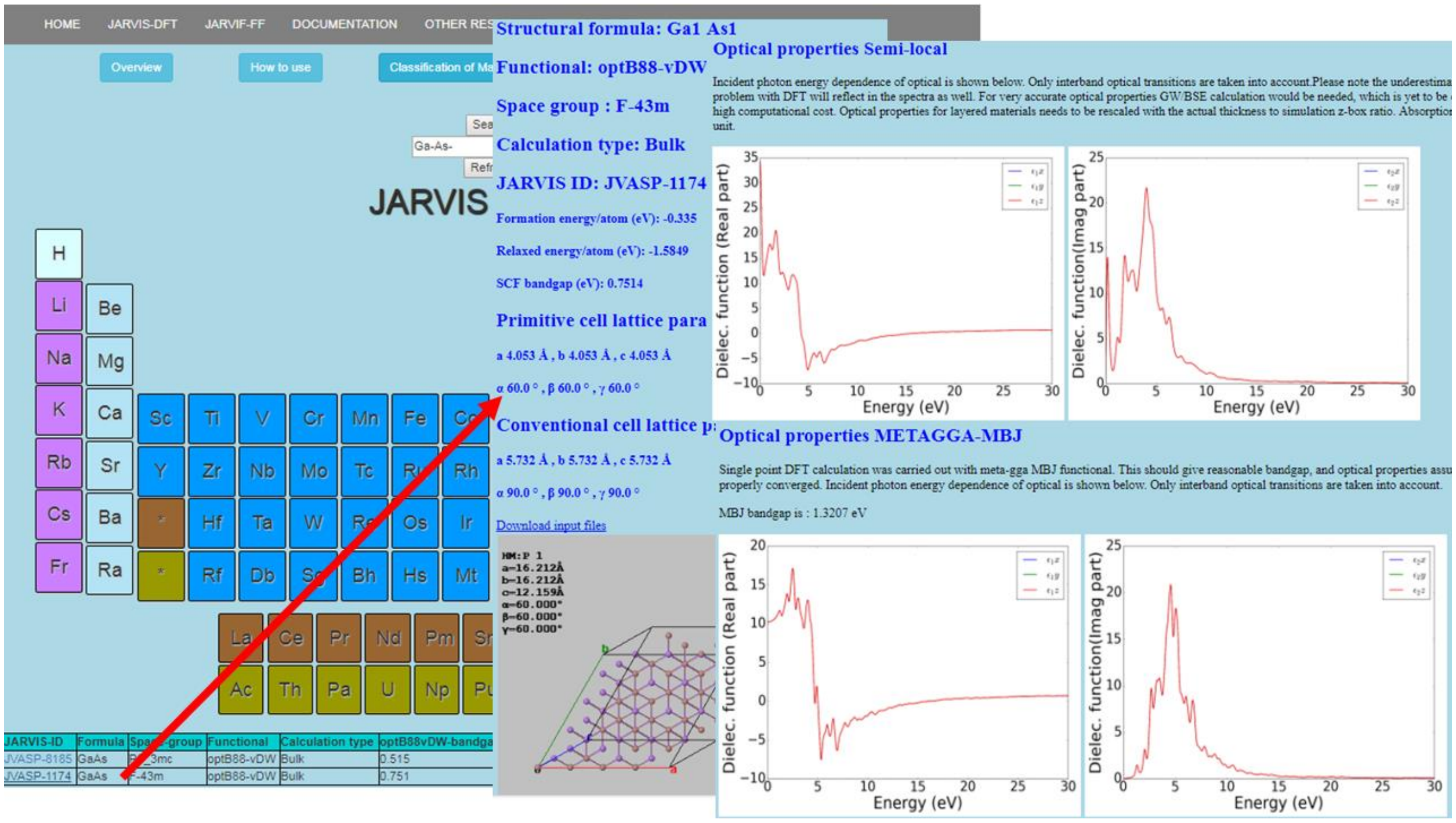

Fig. 2 Snapshot of JARVIS-DFT website.

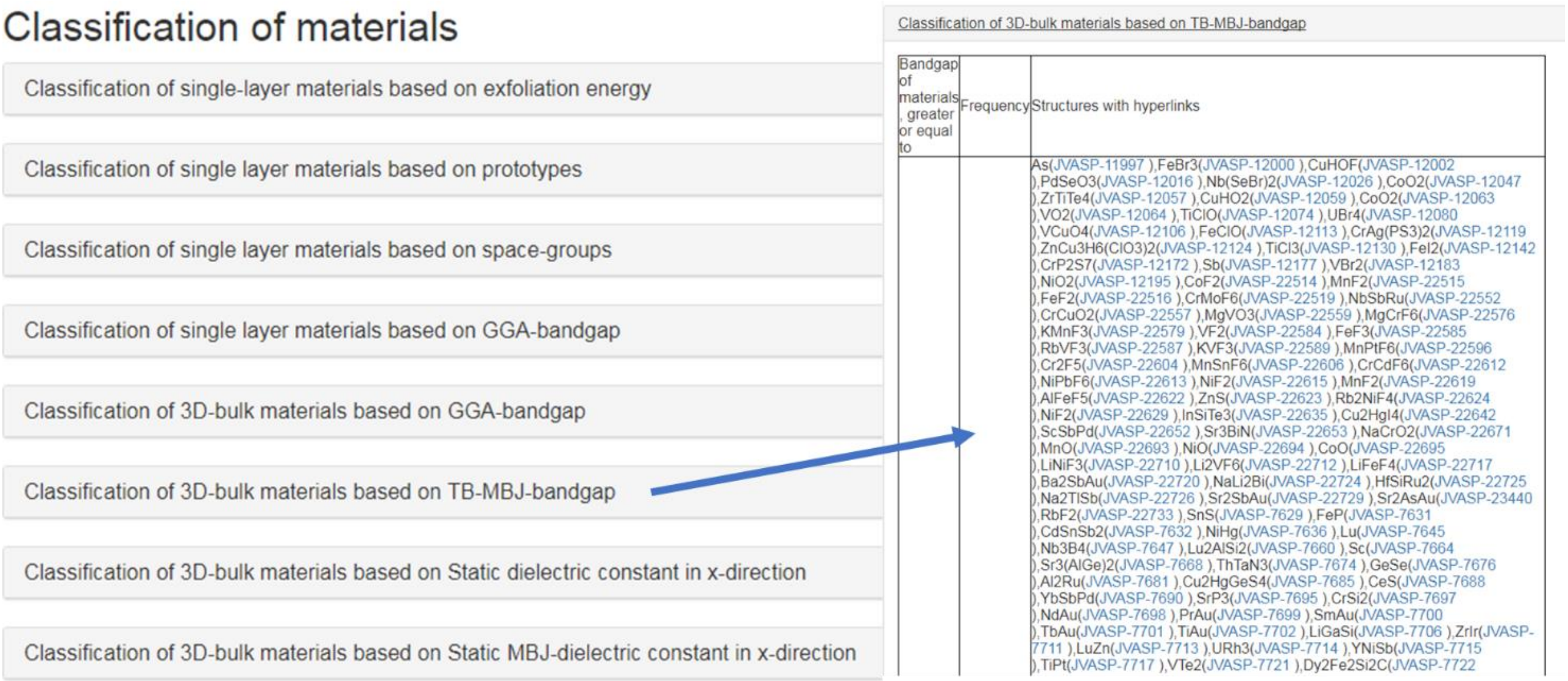

Fig. 3 Material classifications using OPT and MBJ.

**Code availability:**

The code used in this work is provided at https://github.com/usnistgov/jarvis . There are two main scripts in this folder- 1) joptb88vdw.py and 2) master.py. The joptb88vdw.py script heavily utilizes the Pymatgen[8] and ASE[42] codes for file and data management. The joptb88vdw.py generates a series of folders and JSON files starting with keyword ‘ENCUT’ and ‘KPOINT’ denoting the convergence test. An example of an actual calculation is also provided in the folder. After the convergence, the script carries out main geometric relaxation, band structure, optical property with OPT and optical property with MBJ calculations. The master.py takes the argument of the identifier of the database or the structure in ‘VASP’s ‘POSCAR’ format. The master script can tackle both PBS and SLURM formalism used in HPC architecture.

## Data Records

All data computed in this work can be found at https://www.ctcms.nist.gov/~knc6/JVASP.html and https://jarvis.nist.gov/ . A JSON file is also available in a Figshare repository (Data Citation 1). Key variables for the JSON file are shown in Table 1. They include identifiers, structure, bandgaps and dielectric function information with OPT and MBJ methods. The dielectric function data in xx, yy, zz, xy, yz, and zx directions can be used for analyzing the anisotropic nature of the dielectric function. The opt_gap and mbj_gap data can be used to analyze the effect of DFT methodologies on bandgap of a material, where available. The 'jid','mpid'and 'cif' mentioned in Table.1 belong to string-type, while 'opt_gap' and 'mbj_gap' belong to float-type data. The 'mpid' facilitates easy linking to the Materials-project database. Other values such as 'opt_en', 'mbj_en', 'opt_realxx','opt_imagxx', 'mbj_realxx' and 'mbj_imagxx' are arrays with float-type values. The 'real' part in these keys corresponds to real part of dielectric function while 'imag' corresponds to imaginary part of dielectric function in the respective directions. The Pymatgen code can be used to process the 'cif' string-type data. The key 'opt_en' has the same array-size as that of dielectric function data with OPT such as 'opt_realxx', 'opt_imagxx', while 'mbj_en' has the same array-size as that of dielectric function data with MBJ such as 'mbj_realxx' and 'mbj_imagxx'. Packages such as Matplotlib and Gnuplot can be used to plot these arrays and visualize the data. We provide a few examples to explore the JSON files at the github page https://github.com/usnistgov/jarvis/tree/master/jarvis/db/static .

Table. 1 JSON keys for metadata and their descriptions

| **Key** | **Description** |
|---|---|
| **Jid** | JARVIS-DFT calculation identifier |
| **Mpid** | Materials-Project identifier |
| **Cif** | Crystal structure in Crystallographic Information File (CIF) format |
| **opt_gap** | bandgap (unit eV) with OPT |
| **mbj_gap** | bandgap (unit eV) with MBJ |
| **opt_en** | Energy grid array for dielectric function using OPT |
| **opt_realxx, opt_realyy, opt_realzz, opt_realxy, opt_realyz, opt_realzx** | Energy dependent real part of dielectric function in xx, yy, zz, xy, yz and zx directions using OPT |
| **opt_imagxx, opt_imagyy, opt_imagzz, opt_imagxy, opt_imagyz, opt_imagzx** | Energy-dependent imaginary part of dielectric function in xx, yy, zz, xy, yz and zx directions using OPT |
| **mbj_en** | Energy grid array for dielectric function using MBJ |
| **mbj_realxx, mbj_realyy, mbj_realzz, mbj_realxy, mbj_realyz, mbj_realzx** | Energy-dependent real part of dielectric function in xx, yy, zz, xy, yz and zx directions using MBJ |
| **mbj_imagxx, mbj_imagyy, mbj_imagzz, mbj_imagxy, mbj_imagyz, mbj_imagzx** | Energy-dependent imaginary part of dielectric function in xx, yy, zz, xy, yz and zx directions using MBJ |

**Technical Validation:**

As discussed in the method section, the crystal structures were obtained from the Materials Project, which uses PBE for structure optimization. We re-optimize the MP crystal structures with the OPT functional. Most of the MP crystal-structures have Inorganic Crystal Structure Database (ICSD)

IDs, which can be used to obtain experimental lattice parameter information. Hence, we compute PBE and OPT based mean absolute error (MAE) and root-mean-squared error (RMSE) of all the available structures in our database. There are presently 10,052 structures with ICSD IDs in our database. We further classify these structures into predicted vdW and predicted non-vdW structures. We use the lattice-constant criteria[30] and data-mining approaches[31] to identify vdW structures. All the remaining structures are treated as non-vdW bonded. The predicted vdW bonded materials can have vdW bonding in one, two or three crystallographic directions. It is to be noted that exfoliation energy is calculated to predict vdW bonded in materials[30], but the two heuristic methods mentioned above can act as pre-screening criteria for determining vdW bonded structures. Out of 10,052 structures, 2,241 were predicted to be vdW bonded. In addition to the overall MAE and RMSE, we also calculate the same for these two classes of materials as shown in Table 2. As evident from Table. 2, the OPT seems to improve lattice constants in *a*, *b*, *c* crystallographic directions compared to PBE. Significant improvement in lattice parameters is observed for predicted vdW materials, especially in *c*-directions. For predicted non-vdW materials, the errors are similar for OPT and PBE, suggesting that OPT can improved lattice constant predictions for vdW materials without much affecting the predictions for non-vdW bonded materials. Our PBE MAE value for all the materials (0.13 Å) are similar to that obtained by Jianmin et al[43] (0.135 Å) for a smaller set of materials.

Table 2. Mean absolute error (MAE) and root-mean-squared error (RMSE) in *a*, *b* and *c* crystallographic directions computed for all materials in our database with respect to experimental data (ICSD data). To facilitate comparison between the functionals, both MAE and RMSE have been computed for all materials, only for predicted vdW bonded materials and only for predicted non-vdW bonded materials, using Material's project PBE and JARVIS-DFT OPT functional.

| | #Mats. | MAE (*a*) | MAE (*b*) | MAE (*c*) | RMSE (*a*) | RMSE (*b*) | RMSE (*c*) |
|---|---|---|---|---|---|---|---|
| **OPT (All)** | 10052 | 0.11 | 0.11 | 0.18 | 0.29 | 0.30 | 0.58 |
| **PBE (All)** | 10052 | 0.13 | 0.14 | 0.23 | 0.30 | 0.29 | 0.61 |
| **OPT (vdW)** | 2241 | 0.20 | 0.21 | 0.44 | 0.44 | 0.44 | 0.99 |
| **PBE (vdW)** | 2241 | 0.26 | 0.29 | 0.62 | 0.45 | 0.51 | 1.09 |
| **OPT (non-vdW)** | 7811 | 0.08 | 0.08 | 0.11 | 0.23 | 0.24 | 0.39 |
| **PBE (non-vdW)** | 7811 | 0.09 | 0.09 | 0.12 | 0.22 | 0.25 | 0.36 |

As a first validation, we compared the MBJ and OPT bandgaps to experimental values, whenever available. Table 3 displays such a comparison for 54 materials and shows the corresponding results from MP, OQMD, and AFLOW (PBE/PBE+U based data). We also provide identifiers across different databases to facilitate comparison. In general, the values of our OPT and MBJ bandgap data are higher than MP's PBE data, with MBJ data being closer to experiments [44,45]. The mean absolute error (MAE) of MBJ with respect to experimental data is 0.51 eV, while that of OPT is 1.33. The OPT has MAE similar to MP, OQMD, and AFLOW because all of them are primarily PBE based calculations. However, significant improvement is shown with MBJ. Similar results for MBJ gaps versus experimental ones were found by Tran and Blaha et al.[18], validating our methodology. We calculate two MAEs for the data: 1) MAE computed with respect to experiment using all available data for each method, 2) MAE computed with respect to experiment using only data for materials that have results available in all three DFT methods. Both of these values are shown in Table. 3. Both of the MAEs are found to show similar results. It is to be noted that our geometric optimization was performed with OPT, which is different from the one used by Tran-Blaha et al.[18] This explains small differences in MBJ gaps found between our work and by them. Due to the inadequacy of experimental data for all the materials, it is intractable to calculate the error for the whole database. Also, some of the experimental bandgaps were averages of multiple experiments.

The MBJ potential is found to be more suitable for large bandgap insulators and can change the energetics of bands in metallic systems also. We found that some of the materials predicted as metallic using PBE are semiconductors using MBJ, such as Ge and GaAs. To better understand the source of error in the bandgap evaluation, we followed the Materials Project (MP) approach (https://www.materialsproject.org/docs/calculations#Accuracy_of_Band_Structures) and determined a "shifted" MAE for our bandgap evaluations. This treatment allows removing the effect of the DFT systematic underestimation of the gap. To do this, we first fitted a linear equation for the OPT and MBJ data with respect to experiment. The slope was found to be 1.17 and 1.44 for MBJ and OPT, respectively. The slope was then used as a scaling parameter to scale-up the OPT and MBJ data. After the data have been shifted, the MAE with respect to experiment was found to be 0.42 for MBJ, 0.69 for OPT, to compare with the MP result of 0.6. We also calculated the Spearman's coefficient (SC), to measure monotonicity in the bandgap data from different methods compared to experiment. High value for SC suggests that the trends are similar to those in the experimental data. The highest value was obtained for HSE06 (0.97), followed by MBJ (0.94) and AFLOW (0.94). Additionally, we compare the computational time taken during HSE06, MBJ and OPT calculations for a few cases. We find that the MBJ takes about an order of magnitude more computational time than OPT, while HSE06 takes an order of magnitude more computational time than MBJ. A comparison table for computational time for calculations is given in supplementary information (Table. S1).

Next, to understand the trends in the whole database, we compared the bandgaps obtained from the OPT and MBJ as shown in Fig.4a. It is to be noted that many of our calculations for OPT and MBJ are still running; we compare data which are common in both OPT and MBJ only. The blue circles show the MBJ bandgaps while the green ones represent the OPT bandgaps. We also plot the experimental results (red dots) for a small subset (from Table. 3) in the Figure 4a. More specifically, we plotted the three types of data (MBJ, OPT and experiment) against the MBJ results. As the MBJ data are plotted against themselves, they produce a straight line along the diagonal of the plot. For a perfect agreement between OPT and MBJ, all the OPT data should lie on the same straight line. However, most of the OPT data is below the straight line, representing an underestimation of the bandgap. Compared to experiments, the MBJ results describe bandgaps much better than the OPT results. This is shown by the fact that up to about 6 eV most of the experimental data lie on the figure diagonal, while the OPT results lie systematically under it.

The relative difference in OPT and MBJ in bandgap is shown in Fig. S1a. The percentage difference in values for OPT and MBJ are calculated as:

$$\Delta = \frac{|y_{MBJ} - y_{OPT}|}{y_{MBJ}} \times 100\ \% \tag{8}$$

To avoid division by very low or zero values, we calculated percentage differences for materials with OPT gap more than 1 eV. The upper bound of the relative changes in bandgap can range from 30 % up to more than 100 %.

Similar to the bandgap data, the static refractive index in *x*, *y* and *z*-directions are also compared for OPT and MBJ. The static refractive index is related to static dielectric function data as $n(0) = \sqrt{\varepsilon_1(0)}$. The static refractive index in *x*, *y* and *z* directions are shown in Fig. 4 b-d. Like the MBJ bandgaps, the MBJ refractive indices are plotted against itself to give a straight line, which can be used for comparison. A subset of OPT and MBJ static dielectric constant data is shown in Table 3 and compared to experiments. The MAE values of OPT and MBJ static dielectric constant in the *x*-direction are 3.2 and 2.6 respectively, showing the overall superiority of MBJ compared to OPT. It is to be noted that only interband transitions and not intraband are accounted for in our calculations, hence Drude-like transitions are not taken into account[37]. It implies that our dielectric function data should be more accurate for high bandgap materials[18]. Also, in cases where OPT predicts metallic behavior while MBJ predicts semiconductor/insulating, the dielectric function and therefore the static refractive index would be different between OPT and MBJ, because Drude like transitions are not captured in present work. As MBJ bandgaps are more reliable than OPT, the MBJ optical data can be considered more accurate than OPT, especially for low bandgap materials. A very high difference (more than 100 %) in OPT and MBJ refractive index was observed for materials such as $ZnCoF_4$ (as clearly seen in Fig. S1b-S1d) because of the very different bandgaps obtained using OPT and MBJ. We also find that the relative differences between OPT and MBJ refractive indexes are much smaller compared to those for bandgaps. Interestingly, while OPT underestimates the bandgaps compared to experiments, the predicted dielectric functions are relatively close to the experimental measurement, especially for high-bandgap materials. It is because our methodology describes inter-band transitions well but is not suitable for intra-band transitions. Lastly, we also observe that the MBJ static refractive index data are generally lower than the OPT data, as noted in Table 4.

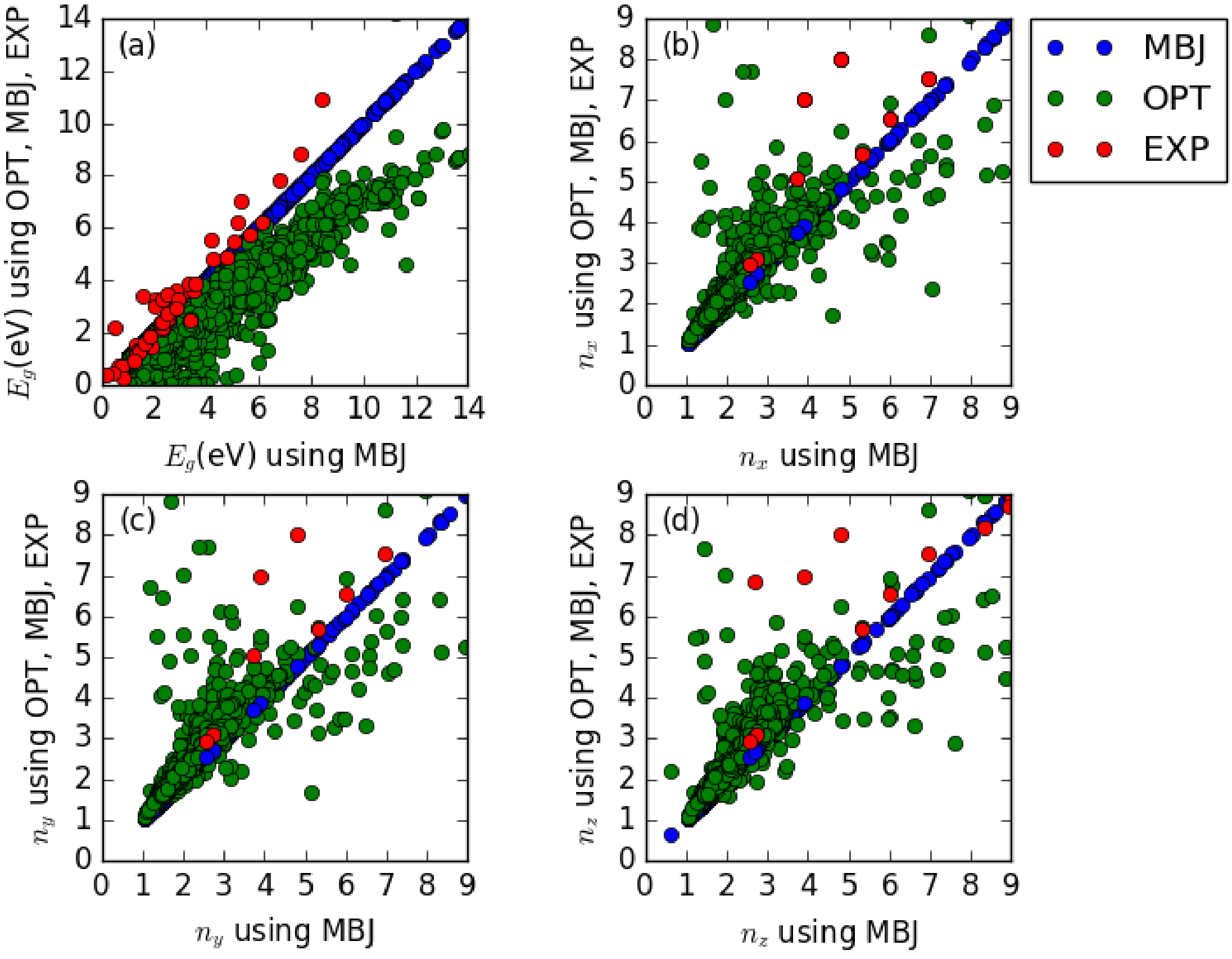

Fig. 4 Comparison of OPT, MBJ, and experimental data. a) fundamental bandgap, b) static refractive index in the *x*-direction, c) static refractive index in the *y*-direction and d) static refractive index in the *z*-direction obtained from OPT and MBJ calculations.

Table 3: Comparison of bandgaps obtained from OPT functional and MBJ potential schemes compared with experimental results and DFT data available in different databases. Materials, space-group (SG), Inorganic Crystal Structure Database (ICSD#) id, Materials-Project (MP#) id, JARVIS-DFT id (JV#), bandgap from MP (MP), bandgap from AFLOW, bandgap from OQMD, our OptB88vdW bandgap (OPT), Tran-Blah modified Becke-Johnson potential bandgap (MBJ), Heyd-Scuseria-Ernzerhof (HSE06) and experimental bandgaps (eV) data are shown. Experimental data were obtained from [18,21,46,47]. MAE denotes the mean absolute error, while SC is the Spearman's coefficient.

| Mats. | SG | ICSD# | MP# | JV# | MP | AFLOW | OQMD | OPT | MBJ | HSE06 | Exp. |
|---|---|---|---|---|---|---|---|---|---|---|---|
| C | Fd-3m | 28857 | 66 | 91 | 4.11 | 4.12 | 4.4 | 4.46 | 5.04 | 5.26 | 5.5 |
| Si | Fd-3m | 29287 | 149 | 1002 | 0.61 | 0.61 | 0.8 | 0.73 | 1.28 | 1.22 | 1.17 |
| Ge | Fd-3m | 41980 | 32 | 890 | 0.0 | 0 | 0.4 | 0.01 | 0.61 | 0.82 | 0.74 |
| BN | $P6_3/mmc$ | 167799 | 984 | 17 | 4.48 | 4.51 | 4.4 | 4.46 | 6.11 | 5.5 | 6.2 |
| AlN | $P6_3mc$ | 31169 | 661 | 39 | 4.06 | 4.06 | 4.5 | 4.47 | 5.20 | 5.49 | 6.19 |
| AlN | F-43m | 67780 | 1700 | 7844 | 3.31 | 3.31 | - | 3.55 | 4.80 | 4.55 | 4.9 |
| GaN | $P6_3mc$ | 34476 | 804 | 30 | 1.74 | 1.91 | 2.1 | 1.94 | 3.08 | 3.15 | 3.5 |
| GaN | F-43m | 157511 | 830 | 8169 | 1.57 | 1.75 | - | 1.79 | 2.9 | 2.85 | 3.28 |
| InN | $P6_3mc$ | 162684 | 22205 | 1180 | 0.0 | 0.0 | - | 0.23 | 0.76 | - | 0.72 |
| BP | F-43m | 29050 | 1479 | 1312 | 1.24 | 1.25 | 1.4 | 1.51 | 1.91 | 1.98 | 2.1 |
| GaP | F-43m | 41676 | 2490 | 1393 | 1.59 | 1.64 | 1.7 | 1.48 | 2.37 | 2.28 | 2.35 |
| AlP | F-43m | 24490 | 1550 | 1327 | 1.63 | 1.63 | 1.7 | 1.79 | 2.56 | 2.30 | 2.50 |
| InP | F-43m | 41443 | 20351 | 1183 | 0.47 | 0.58 | 0.7 | 0.89 | 1.39 | 1.43 | 1.42 |

| Mats. | SG | ICSD# | MP# | JV# | MP | AFLOW | OQMD | OPT | MBJ | HSE | Exp. |
|---|---|---|---|---|---|---|---|---|---|---|---|
| AlSb | F-43m | 24804 | 2624 | 1408 | 1.23 | 1.23 | 1.4 | 1.32 | 1.77 | 1.80 | 1.69 |
| InSb | F-43m | 24519 | 20012 | 1189 | 0.0 | 0.0 | 0.0 | 0.02 | 0.80 | 0.45 | 0.24 |
| GaAs | F-43m | 41674 | 2534 | 1174 | 0.19 | 0.30 | 0.8 | 0.75 | 1.32 | 1.40 | 1.52 |
| InAs | F-43m | 24518 | 20305 | 97 | 0.0 | 0.0 | 0.3 | 0.15 | 0.40 | 0.45 | 0.42 |
| BAs | F-43m | 43871 | 10044 | 7630 | 1.2 | 1.2 | 1.4 | 1.42 | 1.93 | 1.86 | 1.46 |
| $MoS_2$ | $P6_3/mmc$ | 24000 | 2815 | 54 | 1.23 | 1.25 | 1.3 | 0.92 | 1.33 | 1.49 | 1.29 |
| $MoSe_2$ | $P6_3/mmc$ | 49800 | 1634 | 57 | 1.42 | 1.03 | 1.0 | 0.91 | 1.32 | 1.40 | 1.11 |
| $WS_2$ | $P6_3/mmc$ | 56014 | 224 | 72 | 1.56 | 1.29 | 1.4 | 0.72 | 1.51 | 1.6 | 1.38 |
| $WSe_2$ | $P6_3/mmc$ | 40752 | 1821 | 75 | 1.45 | 1.22 | 1.2 | 1.05 | 1.44 | 1.52 | 1.23 |
| $Al_2O_3$ | R-3c | 600672 | 1143 | 32 | 5.85 | 5.85 | 6.3 | 6.43 | 7.57 | 8.34 | 8.8 |
| CdTe | F-43m | 31844 | 406 | 23 | 0.59 | 0.71 | 1.1 | 0.83 | 1.64 | 1.79 | 1.61 |
| SnTe | Fm-3m | 52489 | 1883 | 7860 | 0.04 | 0.25 | 0.3 | 0.04 | 0.16 | 0.17 | 0.36 |
| SnSe | Pnma | 60933 | 691 | 299 | 0.52 | - | 0.6 | 0.71 | 1.25 | 0.89 | 0.90 |
| MgO | Fm-3m | 9863 | 1265 | 116 | 4.45 | 4.47 | 5.3 | 5.13 | 6.80 | 7.13 | 7.83 |
| CaO | Fm-3m | 26959 | 2605 | 1405 | 3.63 | 3.64 | 3.8 | 3.74 | 5.29 | 5.35 | 7.0 |
| CdS | P6_3mc | 31074 | 672 | 95 | 1.2 | 1.25 | 1.4 | 1.06 | 2.61 | - | 2.5 |
| CdS | F-43m | 29278 | 2469 | 8003 | 1.05 | 1.19 | 1.4 | 0.99 | 2.52 | 2.14 | 2.50 |
| CdSe | F-43m | 41528 | 2691 | 1192 | 0.51 | 0.64 | 1.0 | 0.79 | 1.84 | 1.52 | 1.85 |
| MgS | F-43m | 159401 | 1315 | 1300 | 2.76 | 3.39 | 3.6 | 2.95 | 4.26 | 4.66 | 4.78 |
| MgSe | Fm-3m | 53946 | 10760 | 7678 | 1.77 | 1.77 | 1.8 | 2.12 | 3.37 | 2.74 | 2.47 |

| Mats. | SG | ICSD# | MP# | JV# | MP | AFLOW | OQMD | OPT | MBJ | HSE | Exp. |
|---|---|---|---|---|---|---|---|---|---|---|---|
| MgTe | F-43m | 159402 | 13033 | 7762 | 2.32 | 2.32 | 2.5 | 2.49 | 3.49 | 3.39 | 3.60 |
| BaS | Fm-3m | 30240 | 1500 | 1315 | 2.15 | 2.15 | 2.4 | 2.15 | 3.23 | 3.11 | 3.88 |
| BaSe | Fm-3m | 43655 | 1253 | 1294 | 1.95 | 1.95 | 2.9 | 1.97 | 2.85 | 2.79 | 3.58 |
| BaTe | Fm-3m | 29152 | 1000 | 1267 | 1.59 | 1.59 | 1.7 | 1.61 | 2.15 | 2.31 | 3.08 |
| $TiO_2$ | $P4_2/mnm$ | 9161 | 2657 | 5 | 1.78 | 2.26 | 1.8 | 1.77 | 2.07 | 3.34 | 3.30 |
| $TiO_2$ | $I4_1/amd$ | 9852 | 390 | 104 | 2.05 | 2.53 | 2.0 | 2.02 | 2.47 | - | 3.4 |
| $Cu_2O$ | Pn-3m | 26183 | 361 | 1216 | 0.5 | - | 0.8 | 0.13 | 0.49 | 1.98 | 2.17 |
| $CuAlO_2$ | R-3m | 25593 | 3748 | 1453 | 1.8 | 2.0 | 2.4 | 2.06 | 2.06 | - | 3.0 |
| $ZrO_2$ | $P2_1/c$ | 15983 | 2858 | 113 | 3.47 | 3.56 | 4.0 | 3.62 | 4.21 | - | 5.5 |
| $HfO_2$ | $P2_1/c$ | 27313 | 352 | 9147 | 4.02 | 4.02 | 4.5 | 4.12 | 5.66 | - | 5.7 |
| CuCl | F-43m | 23988 | 22914 | 1201 | 0.56 | 1.28 | 0.8 | 0.45 | 1.59 | 2.37 | 3.4 |
| $SrTiO_3$ | Pm-3m | 23076 | 5229 | 8082 | 2.1 | 2.29 | 1.8 | 1.81 | 2.30 | - | 3.3 |
| ZnS | F-43m | 41985 | 10695 | 1702 | 2.02 | 2.67 | 2.4 | 2.09 | 3.59 | 3.30 | 3.84 |
| ZnSe | F-43m | 41527 | 1190 | 96 | 1.17 | 1.70 | 1.5 | 1.22 | 2.63 | 2.37 | 2.82 |
| ZnTe | F-43m | 104196 | 2176 | 1198 | 1.08 | 1.48 | 1.5 | 1.07 | 2.23 | 2.25 | 2.39 |
| SiC | F-43m | 28389 | 8062 | 8158 | 1.37 | 1.37 | 1.5 | 1.62 | 2.31 | - | 2.42 |
| LiF | Fm-3m | 41409 | 1138 | 1130 | 8.72 | 8.75 | 11.0 | 9.48 | 11.2 | - | 14.2 |
| KCl | Fm-3m | 18014 | 23193 | 1145 | 5.03 | 5.05 | 5.3 | 5.33 | 8.41 | 6.53 | 8.50 |
| AgCl | Fm-3m | 56538 | 22922 | 1954 | 0.95 | 1.97 | 1.1 | 0.93 | 2.88 | 2.41 | 3.25 |
| AgBr | Fm-3m | 52246 | 23231 | 8583 | 0.73 | 1.57 | 0.9 | 1.00 | 2.52 | 2.01 | 2.71 |
| AgI | Fm-3m | 52361 | 22919 | 8566 | 0.77 | 1.98 | 1.4 | 0.39 | 2.08 | 2.48 | 2.91 |
| MAE[1] | - | - | - | - | 1.45 | 1.23 | 1.14 | 1.33 | 0.51 | 0.41 | - |
| MAE[2] | - | - | - | - | 1.39 | 1.19 | 1.09 | 1.27 | 0.43 | 0.42 | - |
| S.C. | - | - | - | - | 0.81 | 0.94 | 0.88 | 0.84 | 0.94 | 0.97 | - |

1. MAE calculated with respect to experiment using all available data for each method
2. MAE calculated with respect to experiment using only data for materials that have results available in all three DFT methods.

Next, in Fig. 5 we compare the OPT, MBJ and experimental imaginary part of dielectric function in the *x*-direction for 5a) 1T-$SnSe_2$ ($P\bar{3}m1$), 5b) 2H-$MoS_2$ ($P6_3/mmc$), 5c) Si ($Fd\bar{3}m$), 5d) Ge ($Fd\bar{3}m$), 5e) GaAs ($F\bar{4}3m$) and 5f) InP ($F\bar{4}3m$) . We carried out our experiments for dielectric functional data for 1T-$SnSe_2$ ($P\bar{3}m1$) and 2H-$MoS_2$ ($P6_3/mmc$), while other experimental data were obtained from previous experiments by Aspnes et al[39]. It is clear from Fig. 5 that for MBJ, in general, performs better than OPT peak positions compared to experiments. For $SnSe_2$ and $MoS_2$, both the methodologies give similar result compared to experiments. For 1T-$SnSe_2$, the peaks after 4 eV are more pronounced in DFT than the experiment, which can be attributed to the resolution power of the experiments. In Fig. 5b, the peaks around 2 eV and 4 eV are captured well both in OPT and MBJ for $MoS_2$; however, there is a slight shift in the spectrum due to difference bandgap description between the two functionals at low energy range. We are still investigating the small shift at higher energies, especially for $SnSe_2$. We observe similar spectrum shift due to bandgap underestimation for the cases 5c, 5d, 5e, and 5f. Moreover, peaks at low energy levels using OPT which are absent in MBJ and experimental spectrum. This is likely because when the bandgap is severely underestimated (such as for OPT), the theory predicts inter-band transitions (e.g., valence to conduction band) that simply don't exist because the gap is too high in reality. Such peaks are absent in MBJ based spectrums. It suggests that for low bandgap materials OPT can give unphysical transitions at low energies. However, overall spectrum patterns are similar for OPT and MBJ at higher energies. As observed in Fig. 5, the DFT intensity differs from experiment for some peaks, which can be explained based on 1) the difference in temperature between the experimental setup (generally at room temperature) and the DFT simulation (always at zero Kelvin), and 2) the surface roughness of the sample, which is not included in the calculation. Such differences in peak intensities compared to experiments are also observed for other high-level DFT based methods[48]. In a nutshell, our dielectric function data can be used to complement experimental spectra for instance to allowing to distinguish various peaks. In addition to the peak positions, the DFT data can be used to characterize the orbital nature of the associated electronic transitions, which can provide physical insight into a phenomena[49]. A detailed investigation of all the optical transitions for all the materials will be pursued in future. Other quantities such as refractive index, absorption coefficient, electron-energy-loss spectra (EELS), optical conductivity can be calculated with the dielectric function data. As the dielectric function for materials can be anisotropic, we also provide the dielectric function data in *xx*, *yy*, *zz*, *xy*, *yz*, and *zx* directions, which can be used to calculate frequency dependent birefringence of materials.

Table. 4 Comparison of static dielectric constant for OPT, MBJ and experiment. Experimental data were obtained from [35,47,50-52]

| **Materials** | **SG** | **MP#** | **JV#** | **OPT** | **MBJ** | **Experiment** |
|---|---|---|---|---|---|---|
| $MoS_2$ | $P6_3/mmc$ | 2815 | 54 | $\varepsilon_{11}$ =16.14<br>$\varepsilon_{33}$=9.59 | $\varepsilon_{11}$=15.34<br>$\varepsilon_{33}$=8.99 | $\varepsilon_{11}$=17.0<br>$\varepsilon_{33}$=8.9 |
| $MoSe_2$ | $P6_3/mmc$ | 1634 | 57 | $\varepsilon_{11}$=17.49<br>$\varepsilon_{33}$=10.95 | $\varepsilon_{11}$=16.53<br>$\varepsilon_{33}$=9.71 | $\varepsilon_{11}$=18.0<br>$\varepsilon_{33}$=10.2 |
| $MoTe_2$ | $P6_3/mmc$ | 602 | 60 | $\varepsilon_{11}$=20.73<br>$\varepsilon_{33}$=13.16 | $\varepsilon_{11}$=18.74<br>$\varepsilon_{33}$=11.66 | $\varepsilon_{11}$=20.0<br>$\varepsilon_{33}$=13.0 |
| $WS_2$ | $P6_3/mmc$ | 224 | 72 | $\varepsilon_{11}$=14.59<br>$\varepsilon_{33}$=8.96 | $\varepsilon_{11}$=13.95<br>$\varepsilon_{33}$=8.34 | $\varepsilon_{11}$=11.5<br>$\varepsilon_{33}$=8.2 |
| $WSe_2$ | $P6_3/mmc$ | 1821 | 75 | $\varepsilon_{11}$=15.79<br>$\varepsilon_{33}$=10.2 | $\varepsilon_{11}$=14.32<br>$\varepsilon_{33}$=8.96 | $\varepsilon_{11}$=11.7<br>$\varepsilon_{33}$=8.7 |
| $Al_2O_3$ | R-3c | 1143 | 32 | 3.17 | 2.73 | 3.1 |
| MgO | Fm-3m | 1265 | 116 | 3.1 | 2.54 | 2.95 |
| SiC | $P6_3mc$ | 7631 | 182 | 6.95 | 6.01 | 6.552 |
| C | Fd-3m | 66 | 91 | 5.75 | 5.31 | 5.70 |
| Si | Fd-3m | 149 | 1002 | 13.49 | 10.7 | 11.9 |
| Ge | Fd-3m | 32 | 32 | 27.48 | 15.16 | 16.04 |
| AgI | $P6_3mc$ | 22894 | 8566 | 5.53 | 3.89 | 7.0 |
| AlP | F-43m | 1550 | 1327 | 8.61 | 6.94 | 7.54 |
| BN | $P6_3/mmc$ | 984 | 17 | $\varepsilon_{11}$=4.76 | $\varepsilon_{11}$=3.72 | $\varepsilon_{11}$=5.06 |
| | | | | $\varepsilon_{33}$=3.08 | $\varepsilon_{33}$=2.68 | $\varepsilon_{33}$=6.85 |
| InN | $P6_3mc$ | 22205 | 1180 | 12.22 | 6.8 | 15.3 |
| InP | F-43m | 20351 | 266 | 23.59 | 8.04 | 12.5 |
| BP | F-43m | 1479 | 1312 | 9.1 | 7.94 | 11.0 |
| GaP | F-43m | 2490 | 1393 | 11.59 | 8.33 | 11.11 |
| GaAs | F-43m | 2534 | 1174 | 34.39 | 10.21 | 11.10 |
| InAs | F-43m | 20305 | 97 | 18.13 | 17.95 | 15.15 |
| AlSb | F-43m | 2624 | 1408 | 12.37 | 9.87 | 12.04 |
| GaSb | F-43m | 1156 | 1177 | 22.87 | 13.87 | 15.69 |
| ZnS | F-43m | 10695 | 1702 | 6.24 | 4.8 | 8.0[47] |
| CdTe | F-43m | 406 | 23 | 13.5 | 6.54 | 10.6 |
| HgTe | $P3_121$ | 358 | 8041 | $\varepsilon_{11}$=16.77<br>$\varepsilon_{33}$=22.43 | $\varepsilon_{11}$=11.22<br>$\varepsilon_{33}$ =13.9 | $\varepsilon_{11}$=20<br>$\varepsilon_{33}$=21 |
| $ZnSiP_2$ | I-42d | 4763 | 2376 | $\varepsilon_{11}$=10.95<br>$\varepsilon_{33}$=11.02 | $\varepsilon_{11}$=8.56<br>$\varepsilon_{33}$=8.59 | $\varepsilon_{11}$=11.15<br>$\varepsilon_{33}$=11.7 |
| $ZnGeP_2$ | I-42d | 4524 | 2355 | $\varepsilon_{11}$=13.4<br>$\varepsilon_{33}$=13.6 | $\varepsilon_{11}$=9.02<br>$\varepsilon_{33}$=9.08 | $\varepsilon_{11}$=15<br>$\varepsilon_{33}$=12 |
| $ZnSnAs_2$ | I-42d | 5190 | 8080 | 19.18 | 11.67 | 15.6 |
| MAE($\varepsilon_{11}$) | - | - | - | 3.20 | 2.62 | - |

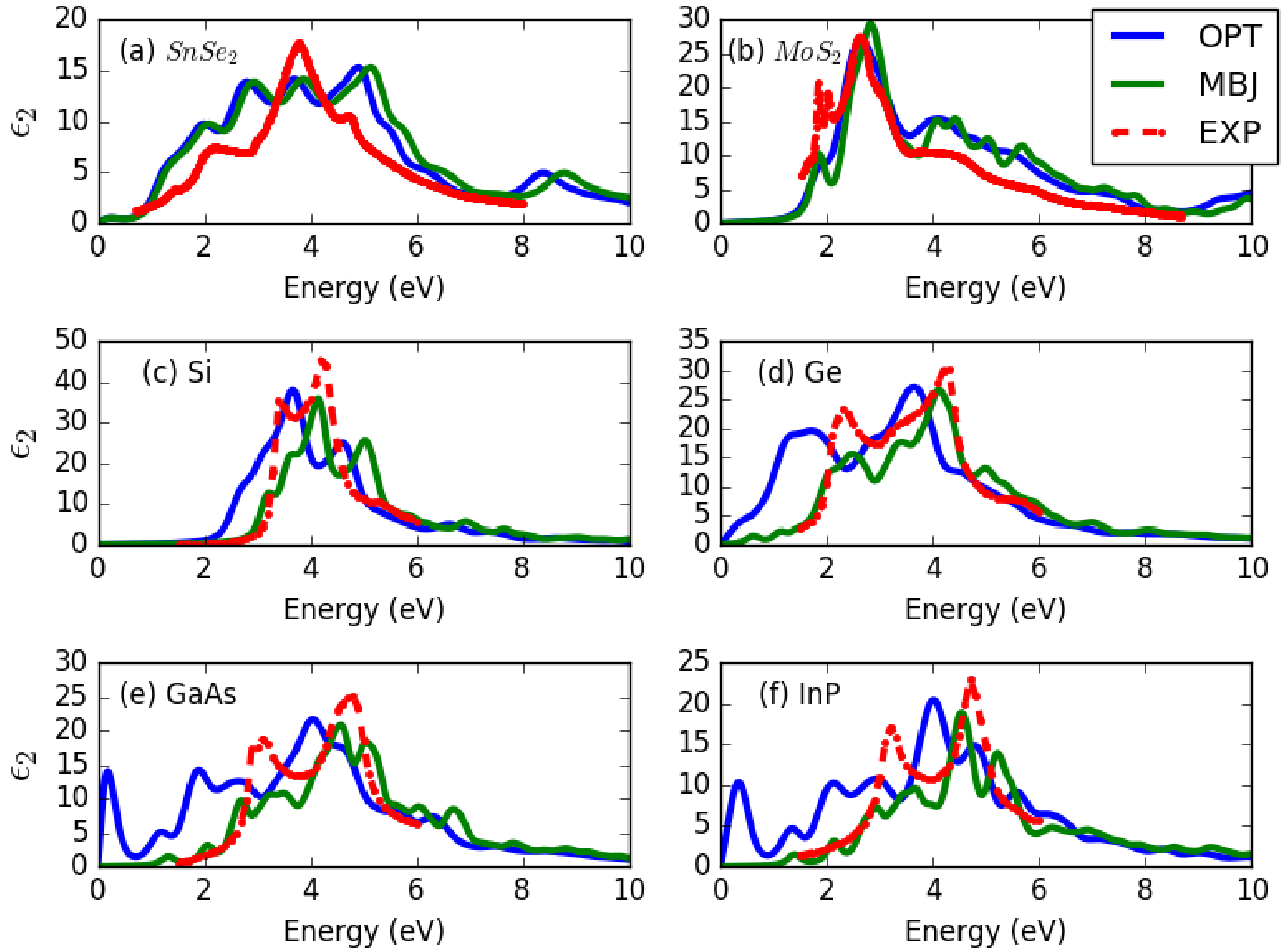

Fig. 5 Imaginary part of the dielectric function in the *x*-direction obtained from OPT, MBJ and experiments. a) 1T-$SnSe_2$ ($P\bar{3}m1$), b) 2H-$MoS_2$ ($P6_3/mmc$), c) Si ($Fd\bar{3}m$), d) Ge ($Fd\bar{3}m$), e) GaAs ($F\bar{4}3m$) and f) InP ($F\bar{4}3m$)

**Usage notes:**

The database presented here represents the largest collection of consistently calculated optoelectronic properties of materials using density functional theory assembled to date. We anticipate that this dataset, and the methods provided for accessing, it will provide a useful tool in fundamental and application-related studies of materials. Our actual experimental verification provides insight into understanding the applicability and limitation of our DFT data.

Based on the list of data, the user will be able to choose particular materials for specific applications. Data mining, data analytics, and artificial-intelligence tools then can be added to guide screening of materials.

**Data Citation:**

1. Choudhary, K. *figshare* https://doi.org/10.6084/m9.figshare.5825994.v1 (2018)

**Acknowledgement:**

We thank Carelyn Campbell, Kevin Garrity, John Vinson and Jason Hattrick-Simpers at NIST for helpful discussions.

**Author contribution:**

K.C. performed DFT calculations, developed the python framework code and webpage, worked on data analysis and verification. Q.Z. and N.V.N. performed ellipsometry experiments. S.C. performed some of the DFT calculations. Z.T. and M.W.N. helped in setting up MDCS MongoDB database for the DFT calculations. F.Y.C. helped in putting up the DFT data in CoRR. A.R. helped in webpage deployment. F.T. assisted in developing the database, designing the convergence criterion, analysis plots and writing the manuscript.

**Competing Interests:**

The authors declare that they have no competing interests.

# References:

1 Polman, A. & Atwater, H. A. Photonic design principles for ultrahigh-efficiency photovoltaics. *Nature materials* **11**, 174-177 (2012).

2 Xiao, Z. *et al.* Efficient perovskite light-emitting diodes featuring nanometre-sized crystallites. *Nature Photonics* (2017).

3 Kawazoe, H., Yasukawa, M., Hyodo, H. & Kurita, M. p-type electrical conduction in transparent thin films of CuAlO2. *Nature* **389**, 939 (1997).

4 Traversa, F. L., Bonani, F., Pershin, Y. V. & Di Ventra, M. Dynamic computing random access memory. *Nanotechnology* **25**, 285201 (2014).

5 Henning, T., Il'In, V., Krivova, N., Michel, B. & Voshchinnikov, N. WWW database of optical constants for astronomy. *Astronomy and Astrophysics Supplement Series* **136**, 405-406 (1999).

6 Forst, C. J., Ashman, C. R., Schwarz, K. & Blochl, P. E. The interface between silicon and a high-k oxide. *Nature* **427**, 53-56 (2004).

7 Brothers, E. N., Izmaylov, A. F., Normand, J. O., Barone, V. & Scuseria, G. E. (AIP, 2008).

8 Ong, S. P. *et al.* Python Materials Genomics (pymatgen): A robust, open-source python library for materials analysis. *Computational Materials Science* **68**, 314-319 (2013).

9 Saal, J. E., Kirklin, S., Aykol, M., Meredig, B. & Wolverton, C. Materials design and discovery with high-throughput density functional theory: the open quantum materials database (OQMD). *Jom* **65**, 1501-1509 (2013).

10 Curtarolo, S. *et al.* AFLOW: an automatic framework for high-throughput materials discovery. *Computational Materials Science* **58**, 218-226 (2012).

11 Perdew, J. P., Burke, K. & Ernzerhof, M. Generalized gradient approximation made simple. *Physical review letters* **77**, 3865 (1996).

12 Petousis, I. *et al.* High-throughput screening of inorganic compounds for the discovery of novel dielectric and optical materials. *Scientific Data* **4**, 160134 (2017).

13 Wang, C. S. & Pickett, W. E. Density-Functional Theory of Excitation Spectra of Semiconductors: Application to Si. *Physical Review Letters* **51**, 597-600 (1983).

14 Chan, M. & Ceder, G. Efficient band gap prediction for solids. *Physical review letters* **105**, 196403 (2010).

15 Yu, L. & Zunger, A. Identification of potential photovoltaic absorbers based on first-principles spectroscopic screening of materials. *Physical review letters* **108**, 068701 (2012).

16 Castelli, I. E. *et al.* New Light-Harvesting Materials Using Accurate and Efficient Bandgap Calculations. *Advanced Energy Materials* **5** (2015).

17 Tran, F. & Blaha, P. Accurate band gaps of semiconductors and insulators with a semilocal exchange-correlation potential. *Physical review letters* **102**, 226401 (2009).

18 Tran, F. & Blaha, P. Importance of the Kinetic Energy Density for Band Gap Calculations in Solids with Density Functional Theory. *J. Phys. Chem. A* **121**, 3318-3325 (2017).

19 Rai, D., Ghimire, M. & Thapa, R. A DFT study of BeX (X= S, Se, Te) semiconductor: modified Becke Johnson (mBJ) potential. *Semiconductors* **48**, 1411-1422 (2014).

20 Setyawan, W., Gaume, R. M., Lam, S., Feigelson, R. S. & Curtarolo, S. High-throughput combinatorial database of electronic band structures for inorganic scintillator materials. *ACS combinatorial science* **13**, 382-390 (2011).

21 Koller, D., Tran, F. & Blaha, P. Merits and limits of the modified Becke-Johnson exchange potential. *Physical Review B* **83**, 195134 (2011).

22 Boujnah, M., Dakir, O., Zaari, H., Benyoussef, A. & El Kenz, A. Optoelectronic response of spinels CdX2O4 with X=(Al, Ga, In) through the modified Becke–Johnson functional. *Journal of Applied Physics* **116**, 123703 (2014).
23 Singh, D. J. Electronic structure calculations with the Tran-Blaha modified Becke-Johnson density functional. *Physical Review B* **82**, 205102 (2010).
24 Feng, W., Xiao, D., Zhang, Y. & Yao, Y. Half-Heusler topological insulators: A first-principles study with the Tran-Blaha modified Becke-Johnson density functional. *Physical Review B* **82**, 235121 (2010).
25 Qiao, J., Kong, X., Hu, Z.-X., Yang, F. & Ji, W. High-mobility transport anisotropy and linear dichroism in few-layer black phosphorus. *Nature communications* **5** (2014).
26 Klimeš, J., Bowler, D. R. & Michaelides, A. Van der Waals density functionals applied to solids. *Physical Review B* **83**, 195131 (2011).
27 Kresse, G. & Furthmüller, J. Efficiency of ab-initio total energy calculations for metals and semiconductors using a plane-wave basis set. *Computational Materials Science* **6**, 15-50, (1996).
28 Kresse, G. & Furthmüller, J. Efficient iterative schemes for ab-initio total-energy calculations using a plane-wave basis set. *Physical Review B* **54**, 11169-11186 (1996).
29 Blöchl, P. E. Projector augmented-wave method. *Physical Review B* **50**, 17953 (1994).
30 Choudhary, K., Kalish, I., Beams, R. & Tavazza, F. High-throughput Identification and Characterization of Two-dimensional Materials using Density functional theory. *Scientific Reports* **7** (2017).
31 Cheon, G. *et al.* Data mining for new two-and one-dimensional weakly bonded solids and lattice-commensurate heterostructures. *Nano Letters* **17**, 1915-1923 (2017).
32 Thonhauser, T. *et al.* Van der Waals density functional: Self-consistent potential and the nature of the van der Waals bond. *Physical Review B* **76**, 125112 (2007).
33 Tawfik, S. A., Gould, T., Stamp, C. & Ford, M. J. Dispersion forces in heterostructures: problem solved? *arXiv preprint arXiv:1712.08327* (2017).
34 Becke, A. & Roussel, M. Exchange holes in inhomogeneous systems: A coordinate-space model. *Physical Review A* **39**, 3761 (1989).
35 Gajdoš, M., Hummer, K., Kresse, G., Furthmüller, J. & Bechstedt, F. Linear optical properties in the projector-augmented wave methodology. *Physical Review B* **73**, 045112 (2006).
36 Moseley, L. & Lukes, T. A simplified derivation of the Kubo-Greenwood formula. *American Journal of Physics* **46**, 676-677 (1978).
37 Wooten, F. *Optical properties of solids*. (Academic press, 2013).
38 Burke, K. The abc of dft. *Department of Chemistry, University of California* (2007).
39 Aspnes, D. E. & Studna, A. Dielectric functions and optical parameters of si, ge, gap, gaas, gasb, inp, inas, and insb from 1.5 to 6.0 ev. *Physical review B* **27**, 985 (1983).
40 Vishwanath, S. *et al.* Controllable growth of layered selenide and telluride heterostructures and superlattices using molecular beam epitaxy. *Journal of Materials Research* **31**, 900-910 (2016).
41 Li, W. *et al.* Broadband optical properties of large-area monolayer CVD molybdenum disulfide. *Physical Review B* **90**, 195434 (2014).
42 Larsen, A. *et al.* The Atomic Simulation Environment—A Python library for working with atoms. *Journal of Physics: Condensed Matter* (2017).
43 Tao, J., Zheng, F., Gebhardt, J., Perdew, J. P. & Rappe, A. M. Screened van der Waals correction to density functional theory for solids. *Physical Review Materials* **1**, 020802 (2017).
44 Nwigboji, I. H. *et al.* Ab-initio computations of electronic and transport properties of wurtzite aluminum nitride (w-AlN). *Materials Chemistry and Physics* **157**, 80-86 (2015).

45 Araujo, R. B., De Almeida, J. & Ferreira Da Silva, A. Electronic properties of III-nitride semiconductors: A first-principles investigation using the Tran-Blaha modified Becke-Johnson potential. *Journal of Applied Physics* **114**, 183702 (2013).
46 Camargo-Martínez, J. & Baquero, R. Performance of the modified Becke-Johnson potential for semiconductors. *Physical Review B* **86**, 195106 (2012).
47 Berger, L. I. *Semiconductor materials*. (CRC press, 1996).
48 Botti, S. *et al.* Long-range contribution to the exchange-correlation kernel of time-dependent density functional theory. *Physical Review B* **69**, 155112 (2004).
49 Choudhary, K. *et al.* Computational discovery of lanthanide doped and Co-doped Y3Al5O12 for optoelectronic applications. *Applied Physics Letters* **107**, 112109 (2015).
50 Kumar, A. & Ahluwalia, P. Tunable dielectric response of transition metals dichalcogenides MX 2 (M= Mo, W; X= S, Se, Te): Effect of quantum confinement. *Physica B: Condensed Matter* **407**, 4627-4634 (2012).
51 Holm, B., Ahuja, R. & Yourdshahyan, Y. Elastic and Optical Properties of α-Al2O3 and k-Al2O3. *Physical Review B*, 777-712 (1999).
52 Yan, J., Jacobsen, K. W. & Thygesen, K. S. Optical properties of bulk semiconductors and graphene/boron nitride: The Bethe-Salpeter equation with derivative discontinuity-corrected density functional energies. *Physical Review B* **86**, 045208 (2012).